\documentclass[twoside]{article}

\usepackage{amsmath,amssymb}

\catcode`\@=11
\long\def\@makefntext#1{
\protect\noindent \hbox to 3.2pt {\hskip-.9pt  
$^{{\eightrm\@thefnmark}}$\hfil}#1\hfill}		

\def\thefootnote{\fnsymbol{footnote}}
\def\@makefnmark{\hbox to 0pt{$^{\@thefnmark}$\hss}}	
	
\def\ps@myheadings{\let\@mkboth\@gobbletwo
\def\@oddhead{\hbox{}
\rightmark\hfil\eightrm\thepage}   
\def\@oddfoot{}\def\@evenhead{\eightrm\thepage\hfil
\leftmark\hbox{}}\def\@evenfoot{}
\def\sectionmark##1{}\def\subsectionmark##1{}}

\oddsidemargin=\evensidemargin
\renewcommand{\thefootnote}{\fnsymbol{footnote}}

\newcounter{sectionc}\newcounter{subsectionc}\newcounter{subsubsectionc}
\renewcommand{\section}[1] {\vspace{12pt}\addtocounter{sectionc}{1} 
\setcounter{subsectionc}{0}\setcounter{subsubsectionc}{0}\noindent 
	{\tenbf\thesectionc. #1}\par\vspace{5pt}}
\renewcommand{\subsection}[1] {\vspace{12pt}\addtocounter{subsectionc}{1} 
	\setcounter{subsubsectionc}{0}\noindent 
	{\bf\thesectionc.\thesubsectionc. {\kern1pt \bfit #1}}\par\vspace{5pt}}
\renewcommand{\subsubsection}[1] {\vspace{12pt}\addtocounter{subsubsectionc}{1}
	\noindent{\tenrm\thesectionc.\thesubsectionc.\thesubsubsectionc.
	{\kern1pt \tenit #1}}\par\vspace{5pt}}
\newcommand{\nonumsection}[1] {\vspace{12pt}\noindent{\tenbf #1}
	\par\vspace{5pt}}

\newcounter{appendixc}
\newcounter{subappendixc}[appendixc]
\newcounter{subsubappendixc}[subappendixc]
\renewcommand{\thesubappendixc}{\Alph{appendixc}.\arabic{subappendixc}}
\renewcommand{\thesubsubappendixc}
	{\Alph{appendixc}.\arabic{subappendixc}.\arabic{subsubappendixc}}

\renewcommand{\appendix}[1] {\vspace{12pt}
        \refstepcounter{appendixc}
        \setcounter{figure}{0}
        \setcounter{table}{0}
        \setcounter{lemma}{0}
        \setcounter{theorem}{0}
        \setcounter{corollary}{0}
        \setcounter{definition}{0}
        \setcounter{equation}{0}
        \renewcommand{\thefigure}{\Alph{appendixc}.\arabic{figure}}
        \renewcommand{\thetable}{\Alph{appendixc}.\arabic{table}}
        \renewcommand{\theappendixc}{\Alph{appendixc}}
        \renewcommand{\thelemma}{\Alph{appendixc}.\arabic{lemma}}
        \renewcommand{\thetheorem}{\Alph{appendixc}.\arabic{theorem}}
        \renewcommand{\thedefinition}{\Alph{appendixc}.\arabic{definition}}
        \renewcommand{\thecorollary}{\Alph{appendixc}.\arabic{corollary}}
        \renewcommand{\theequation}{\Alph{appendixc}.\arabic{equation}}
        \noindent{\tenbf Appendix \theappendixc #1}\par\vspace{5pt}}
\newcommand{\subappendix}[1] {\vspace{12pt}
        \refstepcounter{subappendixc}
        \noindent{\bf Appendix \thesubappendixc. {\kern1pt \bfit #1}}
	\par\vspace{5pt}}
\newcommand{\subsubappendix}[1] {\vspace{12pt}
        \refstepcounter{subsubappendixc}
        \noindent{\rm Appendix \thesubsubappendixc. {\kern1pt \tenit #1}}
	\par\vspace{5pt}}

\newcommand{\textlineskip}{\baselineskip=13pt}
\newcommand{\smalllineskip}{\baselineskip=10pt}

\def\eightcirc{
\begin{picture}(0,0)
\put(4.4,1.8){\circle{6.5}}
\end{picture}}
\def\eightcopyright{\eightcirc\kern2.7pt\hbox{\eightrm c}} 

\newcommand{\copyrightheading}[1]
	{\vspace*{-2.5cm}\smalllineskip{\flushleft
	 }}

\def\abstracts#1#2#3{{
	\centering{\begin{minipage}{4.5in}\baselineskip=10pt\footnotesize
	\parindent=0pt #1\par 
	\parindent=15pt #2\par
	\parindent=15pt #3
	\end{minipage}}\par}} 

\newcommand{\bibit}{\nineit}

\renewenvironment{thebibliography}[1]
	{\frenchspacing
	 \ninerm\baselineskip=11pt
	 \begin{list}{\arabic{enumi}.}
	{\usecounter{enumi}\setlength{\parsep}{0pt}
	 \setlength{\leftmargin 12.7pt}{\rightmargin 0pt} 
	 \setlength{\itemsep}{0pt} \settowidth
	{\labelwidth}{#1.}\sloppy}}{\end{list}}

\newcounter{itemlistc}
\newcounter{romanlistc}
\newcounter{alphlistc}
\newcounter{arabiclistc}

\newcommand{\fcaption}[1]{
        \refstepcounter{figure}
        \setbox\@tempboxa = \hbox{\footnotesize Fig.~\thefigure. #1}
        \ifdim \wd\@tempboxa > 5in
           {\begin{center}
        \parbox{5in}{\footnotesize\smalllineskip Fig.~\thefigure. #1}
            \end{center}}
        \else
             {\begin{center}
             {\footnotesize Fig.~\thefigure. #1}
              \end{center}}
        \fi}

\newcommand{\tcaption}[1]{
        \refstepcounter{table}
        \setbox\@tempboxa = \hbox{\footnotesize Table~\thetable. #1}
        \ifdim \wd\@tempboxa > 5in
           {\begin{center}
        \parbox{5in}{\footnotesize\smalllineskip Table~\thetable. #1}
            \end{center}}
        \else
             {\begin{center}
             {\footnotesize Table~\thetable. #1}
              \end{center}}
        \fi}

\def\@citex[#1]#2{\if@filesw\immediate\write\@auxout
	{\string\citation{#2}}\fi
\def\@citea{}\@cite{\@for\@citeb:=#2\do
	{\@citea\def\@citea{,}\@ifundefined
	{b@\@citeb}{{\bf ?}\@warning
	{Citation `\@citeb' on page \thepage \space undefined}}
	{\csname b@\@citeb\endcsname}}}{#1}}

\newif\if@cghi
\def\cite{\@cghitrue\@ifnextchar [{\@tempswatrue
	\@citex}{\@tempswafalse\@citex[]}}
\def\citelow{\@cghifalse\@ifnextchar [{\@tempswatrue
	\@citex}{\@tempswafalse\@citex[]}}
\def\@cite#1#2{{$\null^{#1}$\if@tempswa\typeout
	{IJCGA warning: optional citation argument 
	ignored: `#2'} \fi}}

\def\pmb#1{\setbox0=\hbox{#1}
	\kern-.025em\copy0\kern-\wd0
	\kern.05em\copy0\kern-\wd0
	\kern-.025em\raise.0433em\box0}

\def\fnt#1#2{\footnotetext{\kern-.3em
	{$^{\mbox{\scriptsize #1}}$}{#2}}}

\def\fpage#1{\begingroup
\voffset=.3in
\thispagestyle{empty}\begin{table}[b]\centerline{\footnotesize #1}
	\end{table}\endgroup}

\def\runninghead#1#2{\pagestyle{myheadings}
\markboth{{\protect\footnotesize\it{\quad #1}}\hfill}
{\hfill{\protect\footnotesize\it{#2\quad}}}}
\font\tenrm=cmr10
\font\tenit=cmti10 
\font\tenbf=cmbx10
\font\bfit=cmbxti10 at 10pt
\font\ninerm=cmr9
\font\nineit=cmti9

\font\eightrm=cmr8

\def\qed{\hbox{${\vcenter{\vbox{			
   \hrule height 0.4pt\hbox{\vrule width 0.4pt height 6pt
   \kern5pt\vrule width 0.4pt}\hrule height 0.4pt}}}$}}

\renewcommand{\thefootnote}{\fnsymbol{footnote}}	

\begin{document}

\runninghead{ Probing the Infrared Structure of Gauge Theories $\ldots$ } { Probing the Infrared Structure of Gauge Theories $\ldots$ }

\normalsize\textlineskip
\thispagestyle{empty}
\setcounter{page}{1}

\copyrightheading{}			

\vspace*{0.88truein}

\fpage{1}
\centerline{\bf 
Probing the Infrared Structure of Gauge Theories:
}
\vspace*{0.035truein}
\centerline{\bf   A Pad\'e-Approximant Approach
             }
\vspace*{0.37truein}
\centerline{\footnotesize  F.A.\ Chishtie, V.\ Elias, 
V.A.\ Miransky
\footnote{On leave  of absence from  Bogolyubov  Institute for Theoretical Physics, 252143, Kiev, Ukraine}
}
\vspace*{0.015truein}
\centerline{\footnotesize\it Department of Applied Mathematics, University of Western Ontario}
\baselineskip=10pt
\centerline{\footnotesize\it London, Ontario, N6A 5B7, Canada}
\vspace*{10pt}
\centerline{\footnotesize T.G. Steele}
\vspace*{0.015truein}
\centerline{\footnotesize\it Department of Physics \& Engineering Physics, University
of Saskatchewan, 116 Science Place}
\baselineskip=10pt
\centerline{\footnotesize\it Saskatoon, Saskatchewan,  S7N 5E2,
Canada}

\vspace*{0.21truein}
\abstracts{
      Pad\'e-approximant treatments of the known terms of the QCD  $\beta$-function are seen to develop
     possible infrared fixed point structure only if the number of fermion flavours is sufficiently
     large.  This flavour threshold is seen to be between six and nine flavours, depending upon
     both the specific choice of approximant as well as on the presently-unknown five-loop  
     $\beta$-function contribution.  Below this flavour threshold, Pad\'e approximants based upon the QCD 
      $\beta$-function manifest the same infrared attractor structure as that which characterizes  the exact
     NSVZ  $\beta$-function of supersymmetric gluodynamics. Such infrared attractor structure is also
     seen to characterize Pad\'e-approximant treatments of vector $SU(N)$ gauge theory in the large $N$  
       limit, suggesting common infrared dynamics for the strong and weak phases of this theory.
}{}{}

\textlineskip			
\vspace*{12pt}			

\noindent
Although the ultraviolet properties of the QCD couplant constant are now quite precisely determined
via knowledge of the four-loop-order $\overline{\rm MS}$ $\beta$-function,\cite{RVL} the infrared behaviour 
of the QCD couplant
remains a mystery. The simplest possibility would be for the existence of an infrared-stable fixed
point (IRFP) in the couplant. However, since QCD with a small number of fermion flavours $n_f$ is a
confining theory, one should anticipate that the relevant degrees of freedom for QCD are essentially
different in the infrared (hadrons) and ultraviolet (quarks and gluons) regions. Such expectations
would argue {\em against} the existence of an IRFP. Indeed, a lattice study has indicated that for three
colours, an infrared fixed point is not possible until $n_f \ge  7$,\cite{Iwasaki} a result 
qualitatively similar to the $n_f$
threshold anticipated from  $\beta$-functions truncated after two-loop order.\cite{BZ}

The present work (which is presented in detail elsewhere\cite{FAC}) utilizes Pad\'e
approximants constructed
from the known terms of the $N_c = 3$ QCD    $\overline{\rm MS}$ $\beta$-function for various 
numbers of flavours in order
to extract infrared properties that are common to all such approximants. Consider for example the
general case of a $k$-loop  $\beta$-function series
$\beta^{(k)} (x) = -\beta_0 x^2 \left( 1 + R_1 x + ... R_{k-1} x^{k-1}
\right)$.
If $k = N + M +1$, this series is sufficient to determine an  $[N\vert M]$-approximant 
 version 
of the same
$\beta$-function:
\begin{equation}
\beta^{[N|M]} (x) = -\beta_0 x^2 \left( {1+a_1 x + ... + a_N
x^N}\right)/\left({1+b_1 x + ... + b_M x^M} \right)
\label{basic_pade}
\end{equation}

\pagebreak
\textheight=7.8truein
\setcounter{footnote}{0}
\renewcommand{\thefootnote}{\alph{footnote}}
\noindent
The $N + M$ coefficients $a_i$ and $b_j$ within the approximant are completely determined 
by the
requirement that the power series expansion of the approximant reproduce the original $k$-loop series. 
Approximants such as that occurring within (\ref{basic_pade}) can be utilized  to obtain information about
whether their underlying function exhibits a zero or a pole, provided such information is not specific
to only  a small subset of possible approximants.\cite{Baker} For example, the approximant in 
(\ref{basic_pade}) can be said
to support the existence of an infrared fixed point provided there exists a positive numerator zero
{\em which precedes any positive zeros of the denominator}.  In ref.\ 4 we present a 
detailed analysis of
possible $[N\vert M]$-approximant  $\beta$-functions ($N + M = 3$) constructed from the known four-loop
 $N_c = 3$  $\beta$-function series. 
 The $[2\vert 1]$- and $[1\vert 2]$-approximant versions of the  $\beta$-function do not 
exhibit a positive numerator zero which precedes
any positive denominator zeros until $n_f\ge   7$ and $n_f\ge 9$, respectively. 
These results clearly support the existence of a threshold in the number of flavours, below which
QCD will not exhibit IRFP infrared content.

To test further the consistency of Pad\'e-approximants in reaching this conclusion, we have also
considered all possible $[N \vert M]$-approximant  $\beta$-functions ($N + M = 4$) constructed from the 
{\em five-loop}
$\beta$-function series 
$\beta(x) = -\beta_0 x^2 \left[1 + \sum_{k=1}^4 R_k x^k\right]$, 
for which we utilize an arbitrary
value for the unknown constant $R_4$ in conjunction with the known values of $R_{1\mbox{--}3}$.\cite{FAC} 
This leads to
$[N\vert M]$-approximant  $\beta$-functions of the form (\ref{basic_pade}) in which the coefficients 
$a_i$ and $b_j$ are themselves
linear functions of $R_4$. We consider $[2\vert 2]$, $[1\vert 3]$ and $[3\vert 1]$ approximants 
(the $[0\vert 4]$ approximant
cannot exhibit a positive numerator zero, and the $[4\vert 0]$ approximant is the truncated series itself),
and examine for a given choice of $n_f$ the range of $R_4$ for which  there exists a positive numerator zero
which precedes any positive denominator zeros. Surprisingly, we find for all three approximants that
{\em no value} for $R_4$ will permit this to happen until $n_f$ equals or exceeds a threshold value. This threshold
value for IRFP's to occur within $[2\vert 2]$-, $[1\vert 3]$-, and $[3\vert 1]$-approximant versions of the 
QCD  $\beta$-function is respectively found to be $n_f = 7$, $n_f = 9$, and $n_f = 6$, values more or 
less consistent with
lattice expectations.\cite{Iwasaki}

In principle, the absence of IRFP's in an approximant-version of a  $\beta$-function can result from either
an absence of positive numerator zeros, or alternatively, from the occurrence of positive denominator
zeros ({\it i.e.,} poles) that precede the occurrence of any positive numerator zeros. In fact, the latter
alternative characterizes all of the approximants considered above except the $[3\vert 1]$-case with $R_4 < 0$.
In the $[2\vert 1]$- and $[1\vert 2]$-approximant versions of the QCD  -function, a  positive approximant pole
is found to precede any approximant zeros when $n_f\le   5$ and $n_f \le  6$, respectively. Moreover, both the 
$[2\vert 2]$- and $[1\vert 3]$-approximant versions of the QCD  $\beta$-function with $R_4$ arbitrary, 
as described above,
are found to have a positive pole which precedes any positive zeros provided $n_f\le   5$, {\em regardless of the
value of} $R_4$. The $[3\vert 1]$-approximant version of the QCD  $\beta$-function has a denominator 
linear in $x$ [the
monomial in the $[3\vert 1]$-approximant denominator is $1 - \left(R_4/R_3\right)x$], 
which necessarily implies that a
positive denominator zero is possible only for one sign (positive) of $R_4$. Nevertheless, if $n_f \le  7$ and 
if $R_4$ {\em is} positive but otherwise arbitrary 
(positivity of $R_4$ for $n_f\le   7$  is anticipated from an asymptotic
Pad\'e-approximant prediction of the $n_f$-dependence within the five-loop contribution to the QCD  
$\beta$-function\cite{EJJ}), the positive zero of the $[3\vert 1]$-approximant denominator is always 
seen to precede any
positive zeros of the $[3\vert 1]$-approximant numerator. Thus, the various approximants considered above
all point toward infrared dynamics for the  $n_f \le  5$ QCD  $\beta$-function that are 
governed by a pole, rather than an IRFP.

It should be noted that such a pole is already known to characterize the exact NSVZ  $\beta$-function of
$N=1$ supersymmetric Yang-Mills theory (supersymmetric gluodynamics),\cite{DRT} and is indicative of
startling infrared dynamics involving a strong and weak phase linked to common infrared properties
({\it e.g.,} equivalent vacuum condensates) and a common infrared limit for the value of the couplant
$\alpha(\mu)$.\cite{KS}  The momentum scale $\mu_c$ characterizing this limiting value is seen to be 
the minimum value
of $\mu$ for which the coupling constant is real. Thus, the value $\mu_c$ determines the boundary of the infrared
region below which supersymmetric gluodynamics as a theory of gluinos and gluons ceases to exist. 
The Pad\'e-approximant analysis described above suggests that such infrared dynamics may also
characterize $n_f\lessapprox 5$ QCD, with $\mu_c$ corresponding to an ${\cal O}(500-600\, {\rm MeV})$ 
infrared boundary on QCD
as a gauge theory of quarks and gluons,\cite{FAC} below which one is compelled to utilize hadronic degrees
of freedom. An infrared pole is also seen to characterize $[2\vert 1]$-, $[1\vert 2]$-, and (for an arbitrary
parametrization of the unknown five-loop contribution) $[2\vert 2]$-, $[1\vert 3]$-, and 
$[3\vert 1]$-Pad\'e-approximant
versions of the  $\beta$-function for gluodynamics in the 't Hooft limit ($N_c\to\infty$).\cite{FAC} 
Infrared dynamics
common to weak and strong phases, as anticipated from such a  $\beta$-function pole, may help account for
the similar glueball spectra obtained via lattice methods\cite{MJT} and via supergravity wave equations in a
black hole geometry,\cite{CC} as found from duality to large $N_c$ gauge theories.

\nonumsection{References}

\end{document}